\documentclass[12pt]{article}
\usepackage[utf8]{inputenc}
\usepackage[english]{babel}
\usepackage[left=0.9in, right=0.9in, top=0.9in, bottom=1.2in]{geometry}
\usepackage[version=3]{mhchem}
\usepackage[pdftex]{graphicx}
\usepackage[colorlinks,urlcolor=blue,citecolor=blue,linkcolor=blue]{hyperref}
\usepackage[usenames,dvipsnames,table,xcdraw]{xcolor}
\usepackage{enumitem}
\usepackage{wrapfig}
\usepackage{overpic}
\usepackage[sort&compress,numbers]{natbib}
\usepackage{hyperref}
\usepackage{nameref}
\usepackage{hyperref}
\usepackage{amsmath}
\usepackage{multirow}
\usepackage{array}
\newcolumntype{L}[1]{>{\raggedright\let\newline\\\arraybackslash\hspace{0pt}}m{#1}}
\newcolumntype{C}[1]{>{\centering\let\newline\\\arraybackslash\hspace{0pt}}m{#1}}
\newcolumntype{R}[1]{>{\raggedleft\let\newline\\\arraybackslash\hspace{0pt}}m{#1}}

\newcommand{\upright}{{}^{|}\!\!\!\to} % for curly arrow
\newcommand{\downright}{{}_{|}\!\!\!\to} % for curly arrow

\usepackage{xcolor}
\usepackage{xspace}

\usepackage[compact]{titlesec}

\usepackage{authblk}

\begin{document}

\newcommand{\HRule}{\rule{\linewidth}{0.5mm}}
\begin{titlepage}
\begin{center}
\HRule \\[0.4cm]
{\huge \bfseries Far-Field Monitoring of Reactor Antineutrinos for Nonproliferation}
\HRule \\[1.5cm]
Viacheslav Li\footnote{E-mail: li68@llnl.gov} for the AIT-WATCHMAN collaboration\\[0.5cm]
{\it \small Lawrence Livermore National Laboratory, Livermore, CA 94550}
%\vspace{2cm}
%\date{\today}
\end{center}

\begin{abstract}
    Numerous experimental efforts have shown that antineutrino-based monitoring provides a non-intrusive means to estimate the fissile content and relative thermal power of nuclear reactors for nonproliferation. However, close proximity to the reactor core is required in order to collect relatively high-statistics data needed for such applications. This has limited the focus of most studies to the so-called ‘near-field’, up to about 200 meters from the reactor core. Until now, there have been no experimental demonstrations dedicated to exploring the nonproliferation potential of large detectors required for long-range monitoring. In this low-statistics regime detailed measurements of the fissile fuel content are not practical, but remote monitoring and discovery of reactors may be achievable. The goal of the Advanced Instrumentation Testbed (AIT) program is to test novel methods for the discovery of reactor cores, specifically in the mid-field to far-field, beyond 200 meters and out to tens or hundreds of kilometers, using kiloton-scale to megaton-scale detectors. The main physical infrastructure of the AIT consists of an underground laboratory in the Boulby mine in Northern England. The site is located at a 26-km standoff from the Hartlepool Reactor Complex, which houses two 1.5-GWth advanced gas-cooled reactors. The first detector to be deployed at the AIT is the WATer CHerenkov Monitor of ANtineutrinos (WATCHMAN). WATCHMAN will use $\sim$6,000 tons of gadolinium doped water in order to detect a few reactor antineutrinos per week from the Hartlepool reactor complex. WATCHMAN will focus on understanding the signal efficiency, radiological backgrounds, and the reactor operational status. Here, the nonproliferation goals are to understand the sensitivity for discovery of one reactor in the presence of another, the discovery of any reactor operations above a well-understood background, and the sensitivity to confirm the declared operational cycles of both reactors. Uniquely, AIT-WATCHMAN also offers a flexible platform at which nascent technologies such as water-based scintillator and fast photomultiplier tubes can be tested in real-world conditions. We present the AIT-WATCHMAN program and status.
\end{abstract}

\vfill
\hrule 
\vspace{0.2cm}
{\footnotesize 
Presented at the 
Institute of Nuclear Materials Management (INMM)
 60th Annual Meeting,\\
 July  14-18, 2019, Palm Desert, California USA.
}

\end{titlepage}

\section{Introduction}

For non-proliferation purposes, it is important to note that even a relatively small reactor of $\sim$10-MWth power could produce a few kg of $^{239}$Pu per year.  There are currently a few hundred active nuclear power reactors and a few dozens under construction~\cite{international2018iaea}. For example, in 2017, about 3/4 of the electricity in France was generated by nuclear power plants, and about 1/5 --- in the US.
The main question is: how to ensure that the nuclear reactors are operated solely for peaceful uses?

\begin{figure}[h]
\centering
\includegraphics[width=0.38\textwidth]{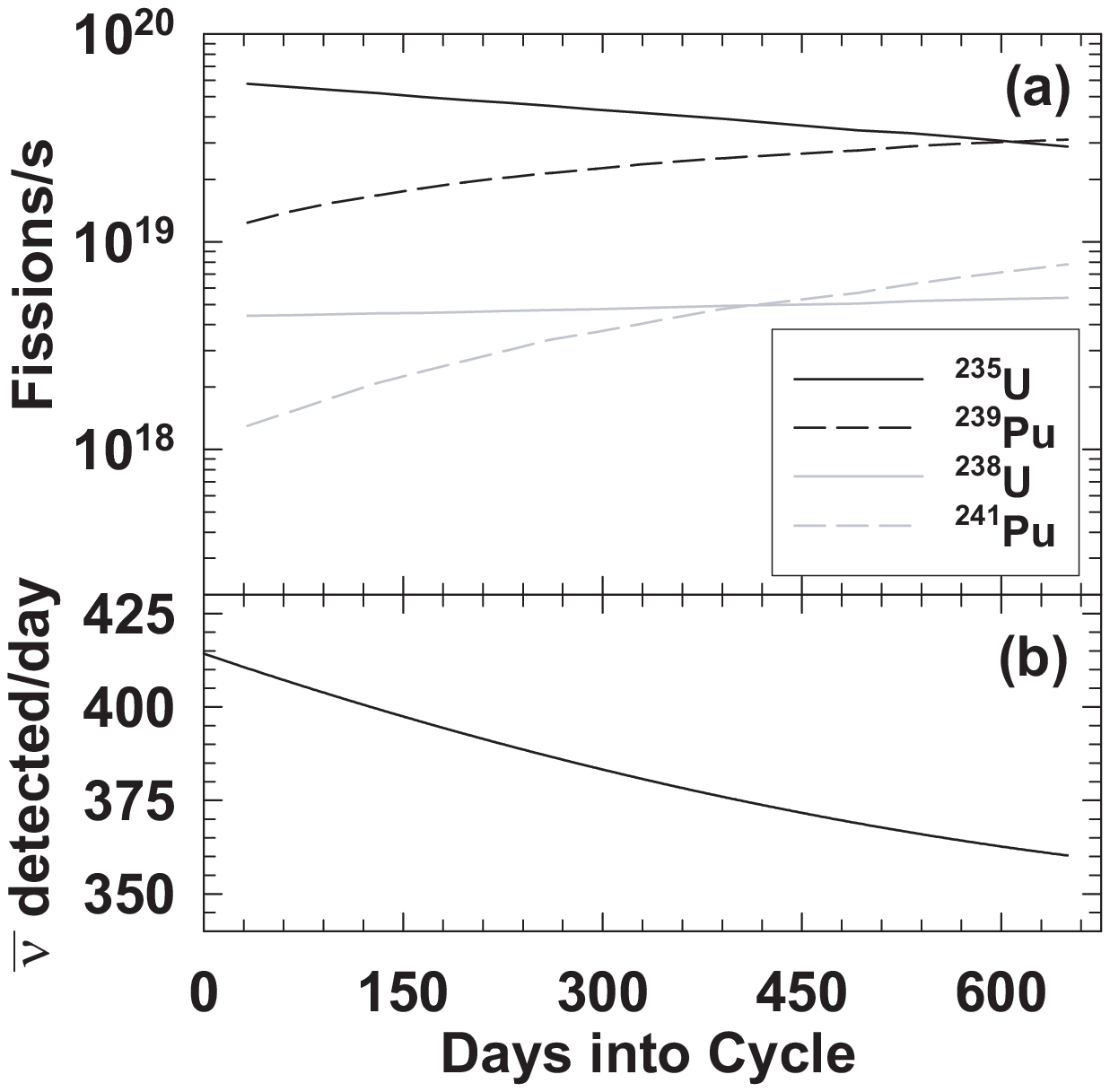}
\hspace{1cm}
\includegraphics[width=0.45\textwidth]{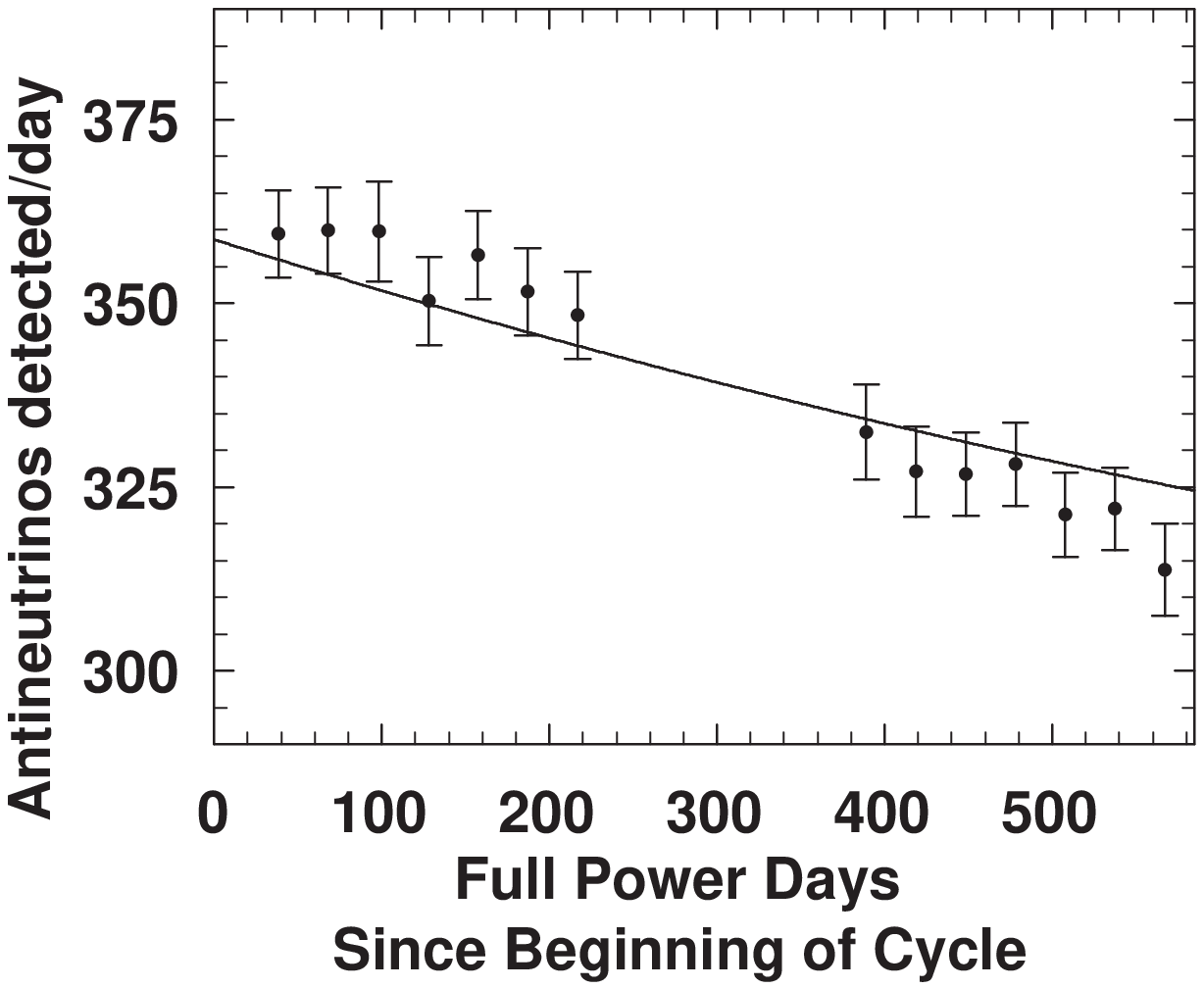}
\caption{Fuel evolution at SONGS predicted and measured based on antineutrino flux,~\cite{Bowden_ev}.}
\label{fig_fuel_evolution}
\end{figure}

\begin{figure}[h]
\centering
\includegraphics[width=0.8\linewidth]{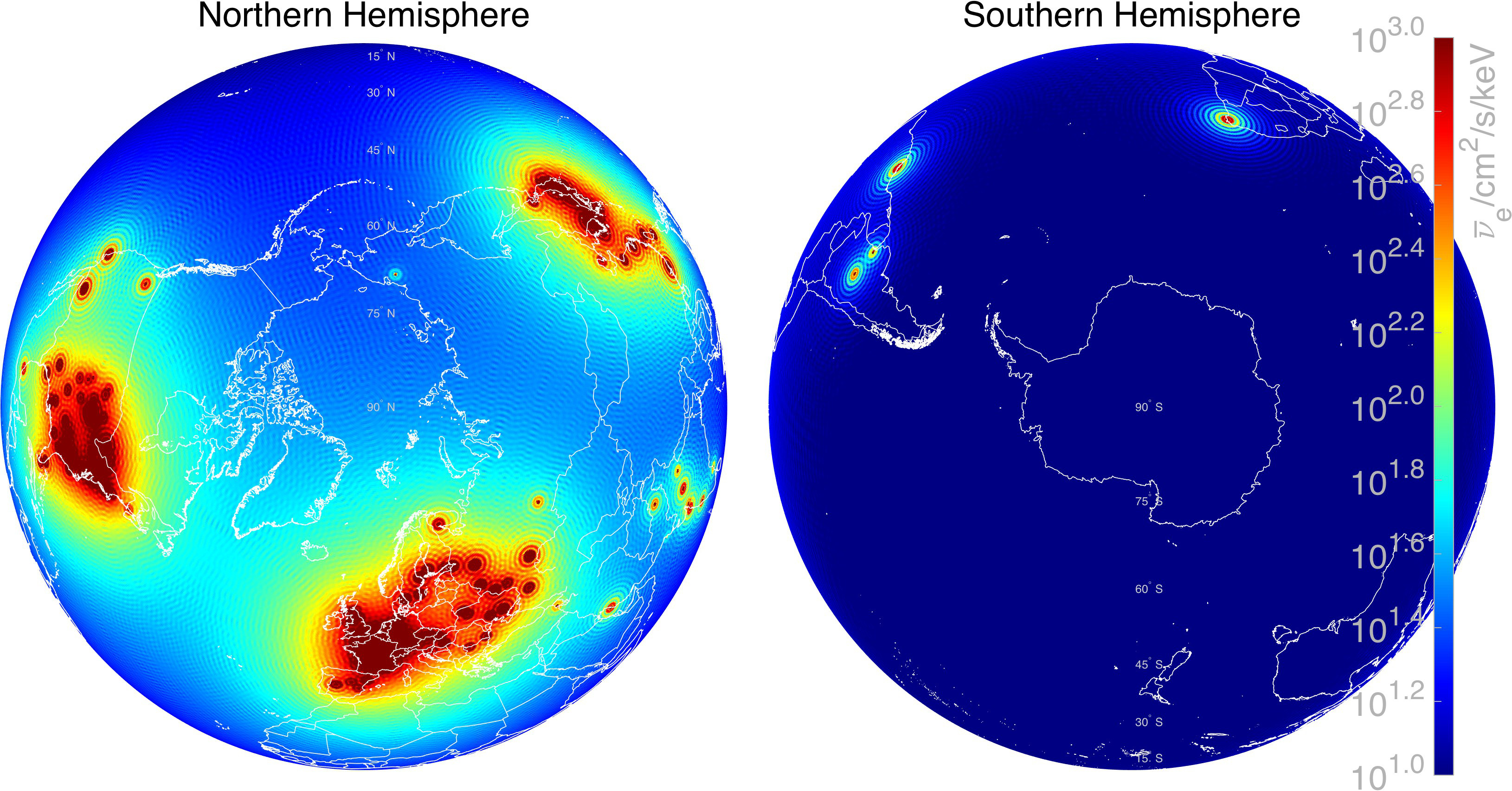} 
\caption{Nuclear reactor neutrino fluxes in Northern and Southern hemisphere. Note the oscillation pattern. Figure is taken from~\cite{Usman:2015yda}.}
\label{fig_AGM_2015}
\end{figure}

On average, a nuclear reactor emits six antineutrinos per fission, each one resulting from a beta decay of a fission daughter. 
The result is $\sim 2 \times 10^{20}$ electron antineutrinos per gigawatt of thermal power.
Antineutrino detection for reactor monitoring purposes was first proposed and realized at the Rovno neutrino laboratory~\cite{Mikaelyan77,Korovkin1984}, including the first measurement of nuclear-fuel burnup using antineutrinos.
Fuel burnup might be a key indicator for tracking a diversion scenario for the production of weapons-grade plutonium under circumstances when direct access to the nuclear-reactor facility is a challenge. Figure~\ref{fig_fuel_evolution} shows the predicted and measured fuel evolution at San Onofre Nuclear Generating Station (SONGS) using antineutrino emissions.
Depending on the location, whether there are many reactors (like in North America, Europe, and East Asia)
or only a couple (Latin America and Africa), shown in Fig.~\ref{fig_AGM_2015}, the backgrounds caused by other reactors vary tremendously and would affect long-range reactor monitoring.

There are many potential diversion scenarios at nuclear reactors, identified by the International Atomic Energy Agency (IAEA)~\cite{international2014iaea}, for which reactor antineutrino monitoring might be applicable.
Good energy resolution for utilizing neutrino oscillations and neutrino directionality are also keys for determining the fuel composition and the geolocation of a reactor~\cite{Lasserre:2010my,Jocher:2013gta}.
However, to draw conclusions about the status of a reactor based on the antineutrino energy spectrum requires an accurate knowledge of the source term. There have only been a couple of experiments, performed at the Institut Laue-Langevin~\cite{VONFEILITZSCH1982162,SCHRECKENBACH1985325,HAHN1989365,Haag:2013raa}, that directly measured the beta spectrum of fission products of $^{235}$U, $^{239}$Pu, and $^{241}$Pu (fissile with slow neutrons) and only recently the beta spectrum originating from $^{238}$U was measured (fissile with fast neutrons). 
The conversion from beta spectrum to antineutrino spectrum is complex, as there are thousands of beta branches, leading to uncertainties in neutrino fluxes~\cite{Mention:2011rk,Huber:2011wv,Hayes:2016qnu} and making the task of precise reactor monitoring more challenging.

There are three reactions by which antineutrinos from nuclear reactors can be detected, presented in Table~\ref{tab_reactions}: inverse beta decay (IBD), elastic neutrino-electron scattering (ES), and coherent elastic neutrino-nucleus scattering (CNNS).
%Neutral current neutrino scattering is coherent for the reactor neutrino energies~\cite{Drukier}.
%Majority of reactor-antineutrino detectors have been based on 
The IBD reaction, a work-horse detection reaction in the majority of reactor experiments, provides a relatively low background delayed-coincidence trigger --- the signal associated with the positron provides a prompt signal, and the neutron capture a delayed signal. 
Other advantages are that the neutrino energy is correlated with the positron energy, while the neutron-capture location can potentially reveal directional information of the antineutrino flux.
A disadvantage of IBD is a relatively high threshold (1.8 MeV), for which only about 1/4 of antineutrinos emitted from a nuclear reactor achieve.
Both ES and CNNS have the advantage that they provide access to the whole reactor antineutrino spectrum, and additionally the ES signal is directional. However, the practical use of ES is limited due to the small cross section and lack of coincidence signal. For these reasons, there are few reactor experiments that have used ES, mostly to evaluate the cross section~\cite{Reines:1976pv}.
Although the CNNS cross section is higher than the IBD, it is difficult to detect the nuclear recoil signal. There are currently few ongoing efforts to utilize CNNS for reactor antineutrino detection (CNNS has been only recently experimentally observed using a spallation neutron source~\cite{Akimov1123}).

\begin{table}[]
    \centering
\begin{tabular}{ |l|c|c|p{5.0cm}|c| l|}
 \hline
 Type &   Reaction & Threshold &  Total cross section (cm$^2$)  & Scaling & Ref. \\
 \hline 
 \hline 
 IBD & $\bar\nu_e  p \xrightarrow{W} e^+  n$ & 1.8 MeV & $\sim 9.53 \times 10^{-44} E_e p_e (1+ \delta)$ & $E^2$ & \cite{Reines1954,Vogel:1999zy,Strumia:2003zx}\\
 
 ES & $\bar\nu_e  e^- \xrightarrow{W,Z} \bar\nu_e e^-$  & 0&  $\sim 0.78 \times 10^{-44} m_e E_\nu$  & $E$ & \cite{Reines:1976pv,HELLFELD2017130}\\
 % Hellfeld paper NIM
 CNNS & $\bar\nu_e A \xrightarrow{Z}  \bar\nu_e A$ &  0 &
 $\sim 0.42 \times 10^{-44} N^2 E_\nu^2$
 %$10^{-41}$~cm$^2$ / $2$~MeV
 & $N^2$, $E^2$ & \cite{Drukier,Akimov1123}\\
 \hline
\end{tabular}
    \caption{Three interactions for reactor antineutrinos.  $E_\nu$ and $E_e$ ($\sim E_\nu - 1.29$~MeV) are the antineutrino and positron total energies respectively. $N$ is the number of neutrons in the nucleus $A$.}
    \label{tab_reactions}
\end{table}

% arxiv
\begin{table}
    \centering
    \begin{tabular}{|L{1.8cm}|C{3cm}||C{1.5cm}|C{1.5cm}||c|c|c|c|c|}\hline
    Fuel isotope & Energy release per fission (MeV) & PWR (start) & PWR (end) & LEU & HEU & PHWR & MOX & GCR\\\hline
  $^{235}$U     & 201.9 & 0.65 & 0.45 & 0.56 & $>.9$     & 0.52 & 0.39  & 0.72 \\\hline
  $^{238}$U     & 205.0 & 0.07 & 0.07 & 0.08 & --        & 0.05 & 0.08  & 0.04 \\\hline
  $^{239}$Pu    & 210.9 & 0.25 & 0.35 & 0.30 & --        & 0.42 & 0.42  & 0.21\\\hline
  $^{241}$Pu    & 213.4 & 0.03 & 0.07 & 0.06 & --        & 0.01 & 0.11  & 0.02\\\hline\hline
  \multicolumn{4}{|l||}{IBD rate per 1\;GW$_\mathrm{th}$ in 1\;ton H$_2$O @50m per day} & 355 & 	387 & 	335 & 	335 & 	363  \\\hline
    \end{tabular}
    \caption{Four main isotopes, {\it the four horsemen}, along with their energy release per fission and average fission fractions in different reactor fuels~\cite{Barna:2015rza}. Note that these fuel fractions change within the cycle. }
    \label{tab_4_isotopes}
\end{table}

Passing through matter mostly unobscured, antineutrinos are an ideal means to monitor nuclear reactor, even from a large distances, since their antineutrino emissions cannot be shielded.
Moreover, the antineutrino rate and spectrum slightly varies depending on the fuel composition. % due to different energy release per fission of different isotopes. 
Recently, there have been a number of studies considering different reactor monitoring scenarios and reactor types, such as: 
future thorium reactors~\cite{Akindele:2015wta}; 
heavy-water reactors~\cite{Christensen:2014pva};
CANDU reactor monitoring%with the SNO+ detector at a $\sim$250-km standoff
~\cite{Baldoncini:2016mbp}; 
experimental light-water LEU reactors~\cite{Carr2019};
and the 
mixed-oxide fuel-burnup effect~\cite{Erickson:2016sdm}. 
%arxiv
The four main reactor-fuel isotopes are presented in Table~\ref{tab_4_isotopes}; the other fissile isotopes, e.g. $^{240}$Pu and $^{242}$Pu, account for less than 0.1\%,~\cite{Djurcic:2008ny}.

Reactor neutrino detectors range greatly in design and target mass, based primarily on the distance between the reactor and the detector. The following (somewhat arbitrary) terminology is applied, to differentiate between different baselines/ranges: {\it near-field} ($\sim$10--200~m), {\it mid-field} ($\sim$200--1,000~m), and {\it far-field} ($\sim$10,000~m). The near-field implies having access as close to the reactor as practicable; and therefore, there must be a cooperative agreement between the reactor management and the detector operator.
The mid-field corresponds approximately to a detector deployment just ``outside the fence''. 
As the number of events detected decreases inversely proportional to the square of the distance, and the neutrino cross sections are minuscule (see Table~\ref{tab_reactions}), most detectors are near-field; with a few mid-field experiments --- PaloVerde, CHOOZ, DoubleCHOOZ, DayaBay, and  RENO~\cite{Boehm:1999gk,Apollonio:2002gd,Abe:2012tg,An:2012eh,RENO:2015ksa}; KamLAND~\cite{Araki:2004mb} has been the only experiment, which successfully detected reactor antineutrinos from an effective $180$-km standoff (averaging from more than 50 reactors in Japan and a few South Korean reactors contributing $\sim$3\% at $\sim$400-km standoff).
Reactor antineutrino experiments played a crucial role in understanding nature of neutrino and the mechanism of neutrino oscillations~\cite{CelebratingNu,Tanabashi:2018oca}.

\section{Advanced Instrumentation Testbed}

\begin{figure}[h]
\centering
\begin{overpic}[height=.5\linewidth]%,right,grid,tics=5]
{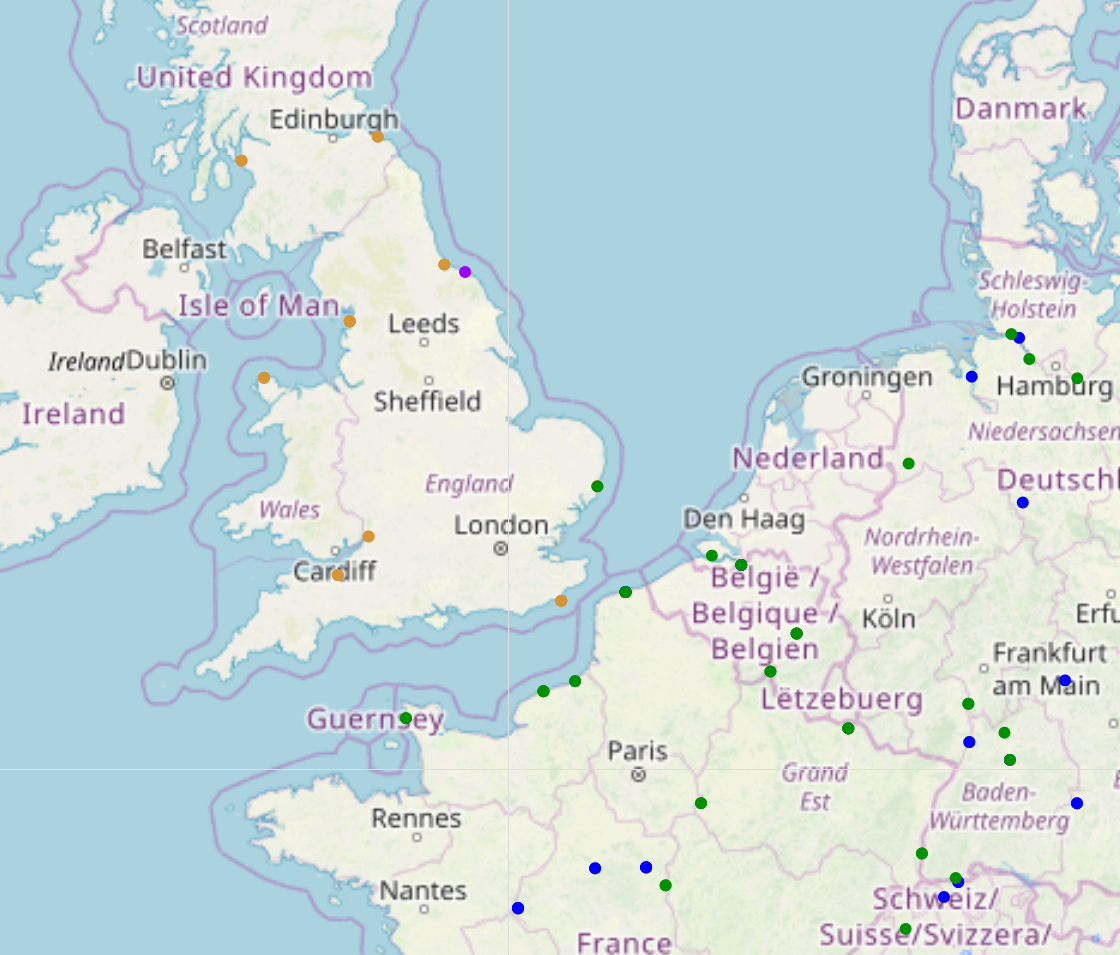}
\put(50,72){\color{red}\vector(-1,-1){10}}
\put(51,72){\color{black}Hartlepool}
\put(50,61.1){\color{red}\vector(-1,0){8}}
\put(51,60){\color{black}WATCHMAN}
\end{overpic}
\hspace{.02\linewidth}
\includegraphics[height=0.5\linewidth]{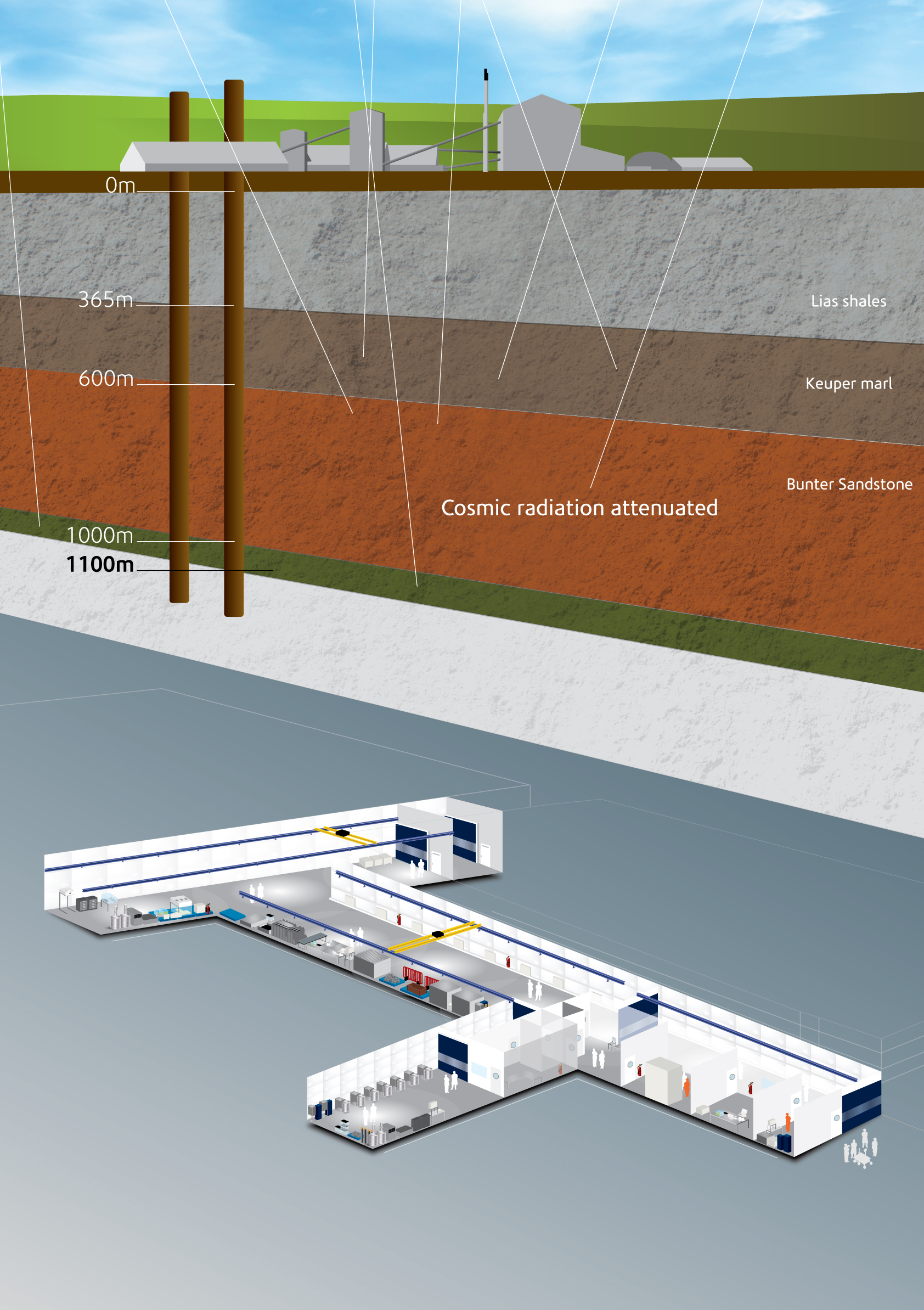}
\caption{{\it Left:} WATCHMAN location relative to British and European reactors. Orange dots --- advanced gas-cooled reactors (most of the UK); blue --- LEU+MOX; green --- LEU pressurized water reactors (most of Europe). Hartlepool is the closest reactor complex to WATCHMAN on the North-East Coast of England. The reactor locations are taken from \href{https://geoneutrinos.org/reactors/}{geoneutrinos.org}, a web-application tool for geo and reactor antineutrinos~\cite{Barna:2015rza}. {\it Right:} Approximate rock profile at the Boulby cite (credit:  STFC Boulby Underground Laboratory).}
\label{fig_map_WM}
\end{figure}

The primary goal of the Advanced Instrumentation Testbed  (AIT) is to demonstrate a reliable long-range reactor monitoring via antineutrinos and to develop new technologies.
AIT program ranges from nuclear non-proliferation studies to scientific --- neutrino oscillations, supernova neutrinos, and geoneutrinos. 
WATCHMAN will be the first detector constructed under the AIT, and will be located at the Boulby Underground Laboratory, $26$-km away from the nearest nuclear reactor, the Hartlepool Nuclear Power Plant on the North-East coast of England, shown in Figure~\ref{fig_map_WM} along with the mine geological profile. 
The main purpose of WATCHMAN is to demonstrate nuclear reactor monitoring of a single reactor in the far-field using a novel detection technology capable of being scaled to megaton-scale volumes.

\begin{figure}[h]
\begin{overpic}[width=1.0\linewidth]%,grid,tics=5]
{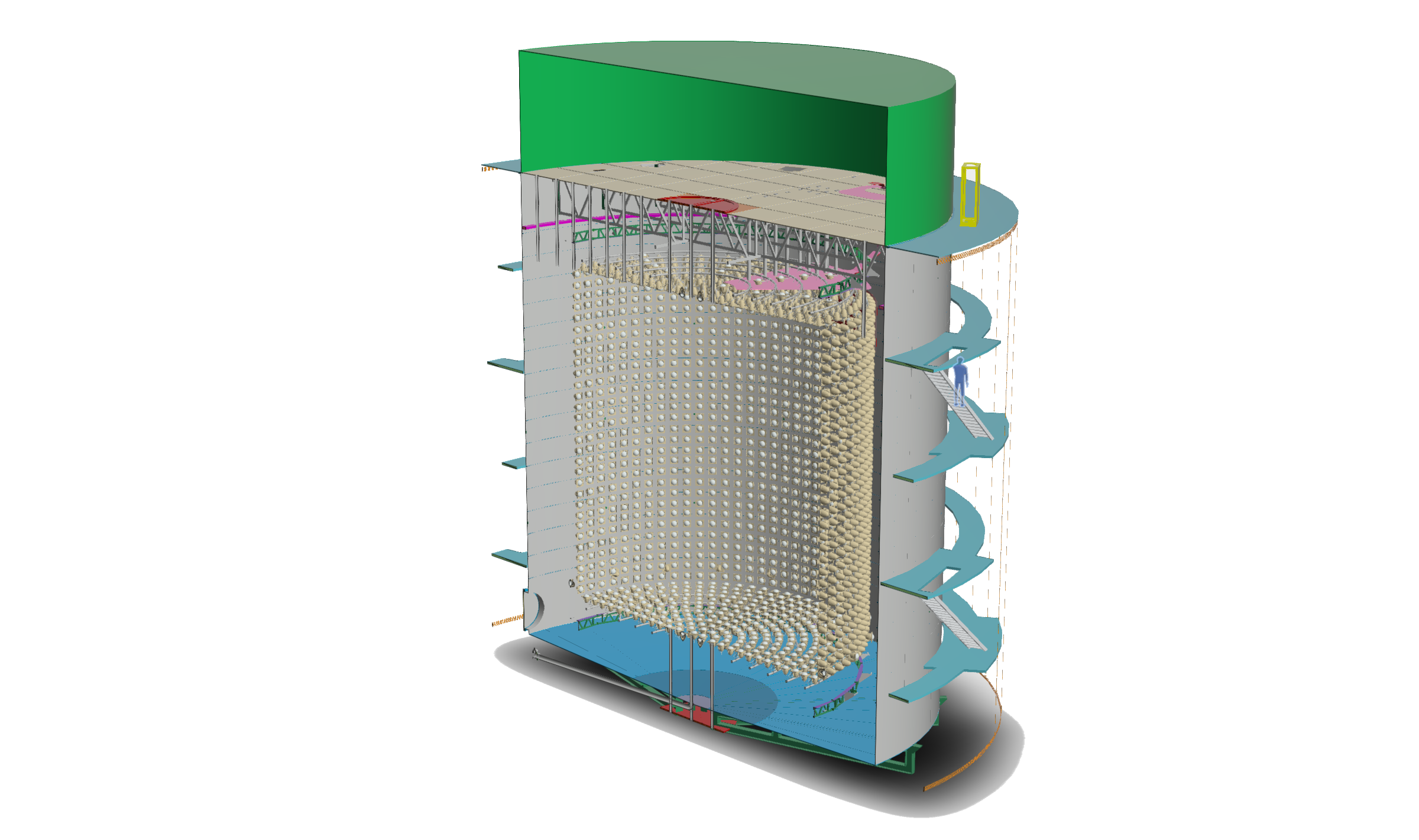}
\put(100,46){\color{red}\vector(-1,0){40}}
\put(81.7,47){\color{black}Calibration ports}
\put(100,38.6){\color{red}\vector(-1,0){41}}
\put(75,39.6){\color{black}PMT support structure}

\put(100,15){\color{red}\vector(-1,0){30}}
\put(77,16){\color{black}External-access levels}

\put(0,52){\color{red}\vector(1,0){49}}
\put(0,53){\color{black}Deck area (DAQ ``huts'' on top)}
\put(0,45){\color{red}\vector(1,0){49}}
\put(0,46){\color{black}Top hatch}
\put(0,35){\color{red}\vector(1,0){37}}
\put(0,36){\color{black}Stainless steel tank}
\put(0,25){\color{red}\vector(1,0){41}}
\put(0,26){\color{black}PMTs (inner and veto)}
\put(0,15){\color{red}\vector(1,0){37}}
\put(0,16){\color{black}Bottom hatch}
\put(0,8.2){\color{red}\vector(1,0){49}}
\put(0,9){\color{black}Emergency drain}
%\put(37,56){\color{blue}\line(0,1){3}}
%\put(62.5,52){\color{blue}\line(0,1){3}}

\put(50,27){\color{blue}\vector(-4,1){13}}
\put(50,27){\color{blue}\vector(4,-1){12}}

\put(49,27){\color{blue}\vector(0,1){18}}
\put(49,27){\color{blue}\vector(0,-1){18}}
\put(46,29){\color{blue}\rotatebox{90}{20 m}}
\put(51,28){\color{blue}\rotatebox{-11}{19 m}}

\put(48.5,27){\color{orange}\vector(0,1){11.5}}
\put(48.5,27){\color{orange}\vector(0,-1){12.5}}
\put(46,20){\color{orange}\rotatebox{90}{13.5 m}}

\put(50,19){\color{orange}\vector(-4,1){9}}
\put(50,19){\color{orange}\vector(4,-1){8}}
\put(50,19.5){\color{orange}\rotatebox{-11}{13.5 m}}

\end{overpic}
\caption{CAD diagram --- vertical slice through the center of the WATCHMAN detector, showing its main components. ``PMTs'' include $\sim$3,600 inner-facing and $\sim$400 outer-facing tubes, with a liner in between --- separating inner and outer (veto) regions. The PMTs are placed about 2 meters from the walls of the stainless steel tank.}
\label{fig_WM_CAD}
\end{figure}

WATCHMAN will be located 1.1~km underground. The detector will be a 19-m-diameter 20-m-height cylindrical stainless steel tank filled with $\sim$6~kt of ultrapure water mixed with gadolinium sulfate, Gd$_2$(SO$_4$)$_3$, with Gd concentration of 0.1\% by weight, i.e. $\sim$6 tonnes of gadolinium. It will comprise an inner detector and a muon veto. The inner detector will be instrumented with about 3,600 PMTs ($\sim$20\% photocoverage) and the veto 400. 
The inner detector fiducial volume (used for the physics analysis) will be a $\sim$1~kt right cylinder. The boundary of the fiducial volume will be located at least 1.5 m from the inner PMTs to protect it from PMT related radioactive background. Figure~\ref{fig_WM_CAD} shows the current placement of main components in the WATCHMAN detector.

The Boulby mine is the deepest operational mine in the UK, and is an active salt and potash mine. 
An extensive network of tunnels (total length over 1,000 km) has been created since operations started in the late 1960s.
Salt, in general, is much more radio pure compared to most other types of rock, which further reduces backgrounds, associated with natural radioactivity of the surrounding rock. 
Most importantly, since the 1990s, the Boulby mine has been hosting the Boulby Underground Laboratory, an internationally-recognized underground scientific laboratory, and has a variety of modern lab space, including clean rooms. The cosmogenic muon flux is reduced by a factor of a million compared to the level at the surface~\cite{Robinson:2003zj}.
The research, performed at the Boulby underground laboratory, spans  multiple disciplines, focusing on 
muon tomography, astrobiology, dark matter, and geology; the AIT program will be adding neutrino physics and non-proliferation research to the list.

% arxiv
\begin{figure}[h]
\centering
\includegraphics[width=1.0\textwidth]{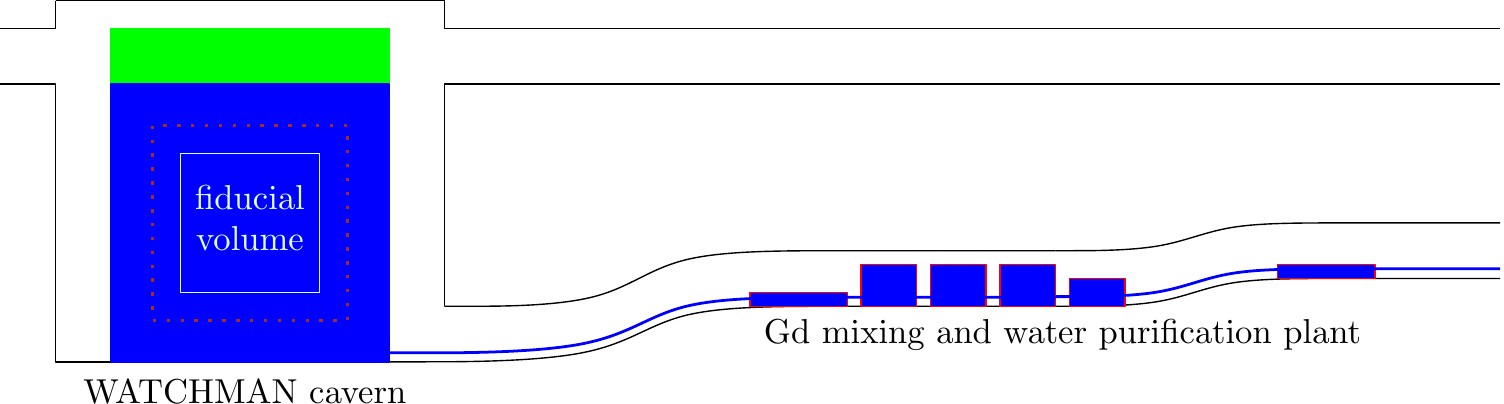}
\caption{WATCHMAN underground layout.}
\label{fig_WM_underground}
\end{figure}

The closest reactor is Hartlepool~\cite{Hartlepool_report} (2 GCR cores, $\sim$3~GW$_{\mathrm{th}}$ max), located at 26~km from Boulby. The next closest reactor is Heysham (4 GCR cores, $\sim$6~GW$_{\mathrm{th}}$ max) --- at 149.4~km from Boulby (contributing $\sim 5 \%$ of the antineutrino flux, relative to the Hartlepool complex); the rest of the world's reactors account for $\sim$10\% of the total flux, which makes Boulby and ideal location to study long-range reactor monitoring from a single reactor complex, since the most of the flux comes from the nearest reactor, Hartlepool.

\begin{figure}[h]
\centering
\includegraphics[width=0.49\textwidth]{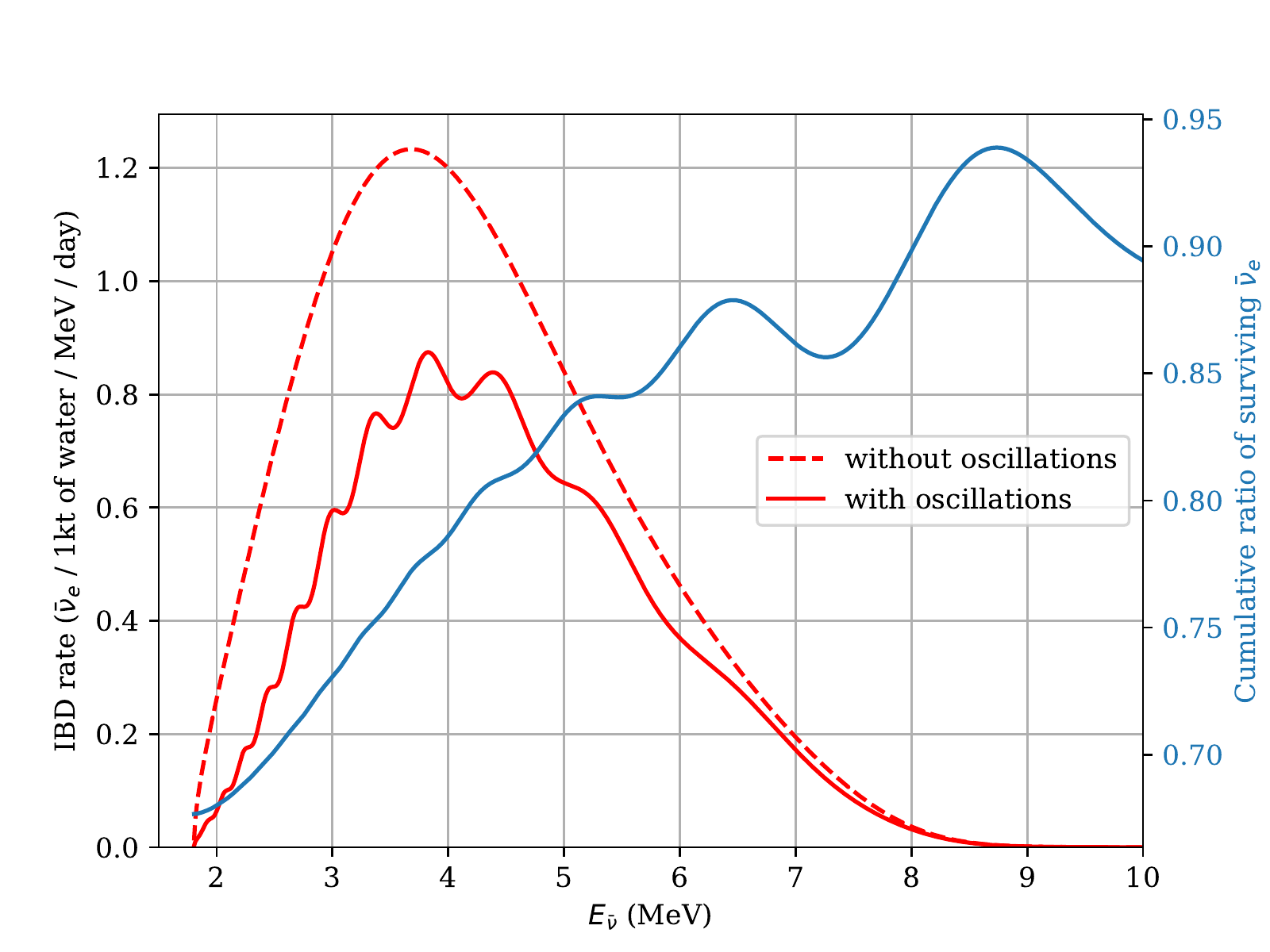}
%\hspace{1cm}
\includegraphics[width=0.49\textwidth]{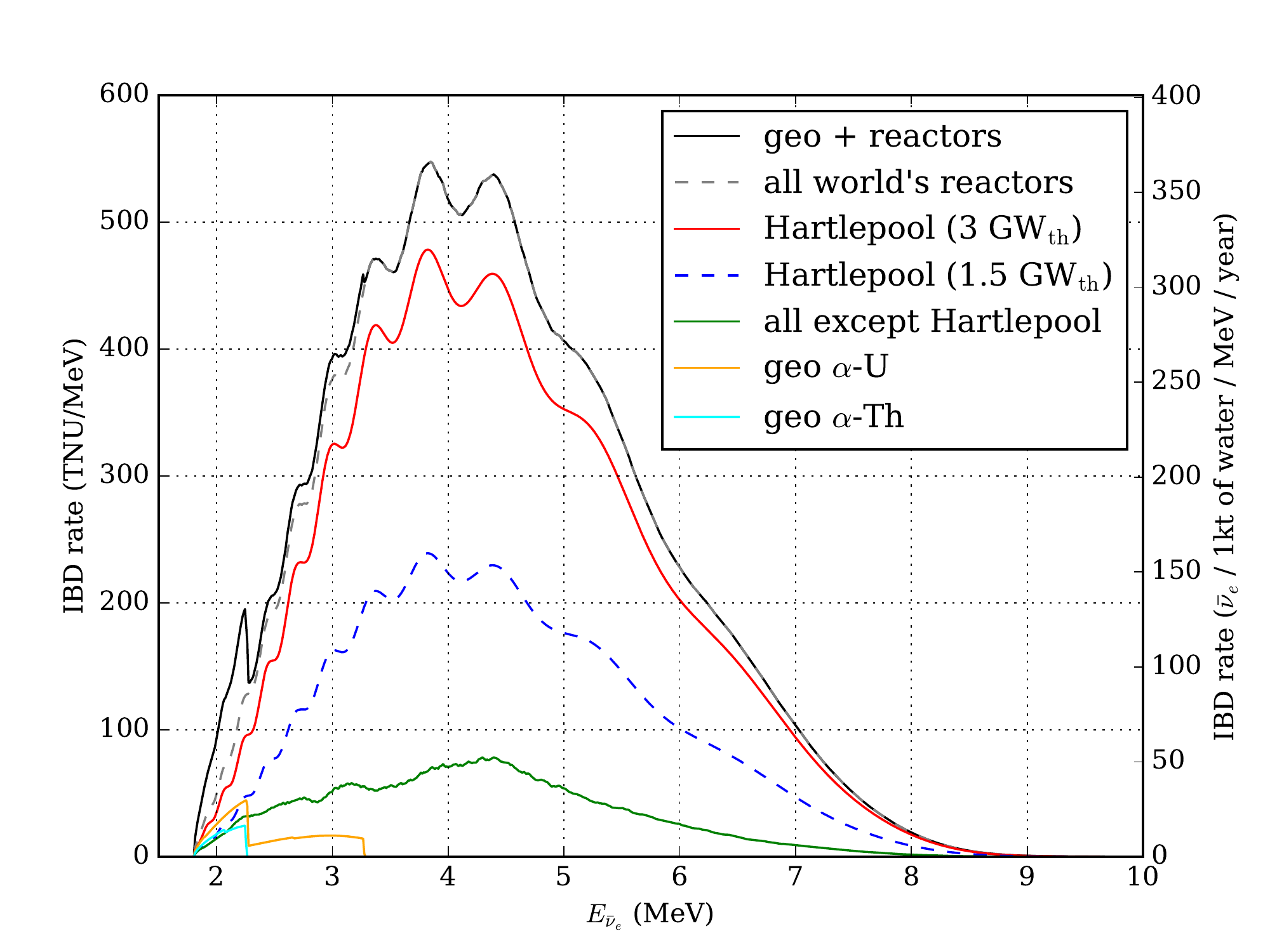}
\caption{{\it Left:} effect of the neutrino oscillations on the Hartlepool flux at Boulby (assuming 3 GWth for this calculation), along with the cumulative survival fraction of antineutrinos.
{\it Right:}
IBD interaction rate at the Boulby underground location, assuming maximum power output for both reactor cores at the Hartlepool complex. The data is taken from \href{https://geoneutrinos.org/reactors/}{geoneutrinos.org}~\cite{Barna:2015rza}.}
\label{fig_IBD_rate}
\end{figure}

The conversion factor for Hartlepool: per each $1$~GWth of the reactor power there is  $\sim 1$ IBD interaction per day in 1~kt of water at Boulby location; Figure~\ref{fig_IBD_rate} represents IBD rate as a function of antineutrino energy. 
In WATCHMAN, an IBD positron produces about 7 photoelectrons per MeV via Cherenkov emission (for scintillator detectors including water-based liquid scintillator, this figure is a couple of orders of magnitude higher); thus, making it almost impossible to detect antineutrinos below $\sim$3--4~MeV. This is the main factor contributing to WATCHMAN detection efficiency, reducing the aforementioned rate to about 40\%. 
The prompt event travels a few centimeters and emits Cherenkov radiation before the positron-electron annihilation (in scintillator, the annihilation 511-keV gammas are visible). The full detection reaction is shown below:

\begin{align*}
 &\ \ \ \downright {\color{red}e^+} + e^- \to  \gamma (511 \ \text{keV}) + \gamma (511 \ \text{keV})  \\
 \bar\nu_e + {}^{1}\mathrm{H} \to \ &  {\color{red}e^+ \mathrm{(prompt: Cherenkov)}}  +   {\color{blue} n  \text{(delayed\ capture:\ Cherenkov\ by\ Compton-recoiled\ }e^-\mathrm{)}} \\
% & \hspace{5cm} \upright {\color{blue}n} + {}^{1}\mathrm{H} \to {}^{2}\mathrm{H} + \gamma's \ \ \ \ \ \ \ \ \ \ \ (Q = 2.2 \ \text{MeV})\\
 & \hspace{5cm} \upright {\color{blue}n} + {}^{155}\mathrm{Gd} \to {}^{156} \mathrm{Gd} + \gamma's\ \ \ (Q = 8.5 \ \text{MeV})\\
 & \hspace{5cm} \upright {\color{blue}n} + {}^{157}\mathrm{Gd} \to {}^{158} \mathrm{Gd} + \gamma's \ \ \ (Q = 7.9 \ \text{MeV})%7937keV
 \end{align*}

Gadolinium has the highest thermal-neutron-capture cross section of any stable element. 
An effective cross section is $\sim$4.9$\times$10$^4$ barns for thermal-neutron capture on Gd, mainly contributing by two isotopes: 
6.09$\times 10^4$ barns for $^{155}$Gd (14.80\% nat. ab.) and 
2.54$\times 10^5$ barns for $^{157}$Gd  (15.65\% nat. ab.).
The 0.1\% Gd-doping by weight corresponds to about 9 neutron captures on Gd per every capture on H. 
In addition to its high neutron-capture cross section,
the Gd(n,$\gamma$) reaction results in a few gammas summing to $\sim$8~MeV, which provides sufficient light output to be detectable in a Cherenkov detector (compared to 2.2~MeV, in the case of neutron capture on hydrogen).
A detailed study of the gamma spectra along with gamma multiplicity in the $^{155,157}$Gd(n,$\gamma$) reactions has recently been carried out~\cite{Hagiwara:2018kmr,Tanaka:2019hti}. 
Adding gadolinium also reduces the capture time needed for delayed coincidence: for the gadolinium concentration in WATCHMAN, the mean thermalization and neutron-capture time is of the order of few tens of microseconds (instead of few hundreds of microseconds, in case of hydrogen).
The prompt and delayed events are also correlated by distance --- typically within $\sim$10~cm.

Gadolinium has been successfully used in liquid-scintillator detectors, and also recently in water detectors (WAND, EGADS). Table~\ref{tab_Gd_projects} represents global research efforts.
Large-volume water-Cherenkov detectors have a great record of reliability and discovery potential since the pioneering IMB and Kamiokande experiments~\cite{Bionta:1987qt,Hirata:1987hu}; which makes water-Cherenkov technology well-established and cost-effective for scalability purposes. 
In 2001, it was proposed to add gadolinium in water to monitor IBD events from fission nuclear explosions at long range\cite{Bernstein2001}. Adding gadolinium makes the detector sensitive to both IBD signals --- the prompt positron and delayed neutron (via capture on Gd). In 2003, J.~Beacom and M.~Vagins~\cite{Beacom:2003nk} separately proposed to add gadolinium to a large-volume water-Cherenkov detector (namely, SuperKamiokande or SuperK) to detect diffuse supernova neutrino background (DSNB, or relic supernovae antineutrinos). In the early 2000s’, as the Bernstein paper notes, the cost of gadolinium was \$100 per gram. Since then the cost plummeted by a few orders of magnitude and gadolinium has become an affordable commodity. 
Although a SuperK scale (50 ktons) is needed to sufficiently register DSNB, for reactor monitoring purposes a detector can be made smaller and placed at an appropriate distance from a reactor. EGADS has successfully demonstrated the technology of doping water with gadolinium on a 200-ton scale.

Similar to EGADS and SuperK-Gd designs, the water system of WATCHMAN will have a water-purification system and Gd-mixing system, schematically shown in Figure~\ref{fig_WM_underground}. 
The technological challenge that has been successfully demonstrated originally for EGADS is to selectively retain gadolinium Gd while removing all other impurities --- a so-called {\it molecular band-pass filtration}, a new method developed by M.~Vagins, based on a staged nanofiltration~\cite{Xu:2016cfv}. Since the gadolinium itself is not water-soluble, it is added in the form of gadolinium sulfate, which is water-solubale. The toxicity of gadolinium sulfate is relatively low and a  method of tracing gadolinium in the environment has been developed~\cite{Ito:2019jhr}.

The construction of WATCHMAN is scheduled to begin in 2020. After two years of operation, WATCHMAN will be refilled with a water-based liquid scintillator (WbLS), a new technology developed by M.~Yeh's group at Brookhaven National Laboratory~\cite{Bignell:2015oqa}, which allows for the detection of an additional small level of scintillation light with a slightly reduced Cherenkov signal.
While more R\&D is needed to accurately determine the light propagation properties of WbLS, the aim is to achieve
higher detection efficiency and better energy and vertex reconstruction.
The Advanced Instrumentation Testbed will be a platform for testing future technologies relevant to  remote reactor antineutrino monitoring and  exploring scalability to large-volume and cost-effective reactor antineutrino monitors.

\begin{table}[]
    \centering
    \begin{tabular}{|l|c|c|r|}\hline
        Project & Mass & Gd content & Ref.  \\\hline\hline
Dazeley et al. & 0.25 ton & 0.2\% wt. GdCl$_3$ & \cite{DAZELEY2009616}\\\hline
Watanabe et al. & 2.4 liters & 0.2\% wt. GdCl$_3$ & \cite{WATANABE2009320}\\\hline
WAND & 1.0 ton & 0.4\% wt. GdCl$_3$ & \cite{DAZELEY201532}\\\hline
WATCHBOY & 2 tons & 0.2\% wt. GdCl$_3$ & \cite{Dazeley:2015uyd}\\\hline
ANNIE & 26 tons & 0.2\% wt. Gd$_2$(SO$_4$)$_3$ & \cite{Back:2017kfo}\\\hline
EGADS & 200 tons & 0.2\% wt. Gd$_2$(SO$_4$)$_3$ & \cite{Xu:2016cfv} \\\hline
WATCHMAN  & 6 kt (1 kt fiducial)  & 0.2\% wt. Gd$_2$(SO$_4$)$_3$ & \cite{Askins:2015bmb} \\\hline
SuperK-Gd & 50 kt (22.5 kt fiducial) &  0.2\% wt. Gd$_2$(SO$_4$)$_3$ & \cite{Beacom:2003nk} \\\hline
    \end{tabular}
    \caption{Global research efforts on Gd-H$_2$O technology.} %Projects that employ Gd-doped water.}
    \label{tab_Gd_projects}
\end{table}

\section{Summary}

Building on proven technology, WATCHMAN will be the first large-scale Gd-doped water-Cherenkov detector with a primary goal of far-field monitoring of a single nuclear reactor complex. On a scientific side, WATCHMAN will be added to a network of supernova neutrino monitors, and has potential to provide additional information for reactor neutrino oscillations at far-field.

In recent years, an extensive simulations and data-analysis software tool set has been developed.
PMT tests and DAQ instrumentation downselect processes have already started. The
main components of the proposed water purification and gadolinium mixing have been identified. 
The cavern excavation is scheduled to begin in 2020.
Over the course of 2 years of operation, we expect to detect $\sim$1,000 IBD events originating from the two nuclear reactor Hartlepool site.
After the necessary tests and the establishment of a way to continuously purify WbLS, a proposal to upgrade the detector to a water-based liquid scintillator will be considered.  
The introduction of WbLS will likely increase light yield, and therefore decrease the energy detection threshold, resulting in potential sensitivity to geoneutrinos.

The AIT-WATCHMAN collaboration has been growing rapidly and now includes almost a hundred members from a few dozen institutions, most of whom have experience in building large-scale neutrino detectors.

\section{Acknowledgement}

We acknowledge the support of the U.S. Department of Energy, the U.K. government, the U.K.  Science and Technology Facilities Council, and the Boulby Underground Science Facility.

This work was performed under the auspices of the U.S. Department of Energy by Lawrence Livermore National Laboratory under Contract DE-AC52-07NA27344. 
Release number LLNL-PROC-775357.

\bibliographystyle{apsrev4-1}

\newpage
\bibliography{refs.bib}

\end{document}